\documentclass[11pt]{article}
\usepackage{authblk}
\usepackage[utf8]{inputenc}
\usepackage[T1]{fontenc}
\usepackage{geometry}
\usepackage{amsmath,amssymb,amsfonts,amsthm,mathtools}
\usepackage{hyperref}
\usepackage{graphicx} % Required for inserting images

\usepackage{times}
\usepackage{subfiles}
\usepackage{array}
\usepackage{adjustbox}
\usepackage{paralist}
\usepackage{wrapfig}
\usepackage{float}

\newtheorem{assumption_}{\textbf{Assumption}}
\newtheorem{lemma_}{\textbf{Lemma}}
\newtheorem{remark_}{\textbf{Remark}}

\newtheorem{theorem_}{\textbf{Theorem}}
\newtheorem{definition_}{\textbf{Definition}}
\newtheorem{proposition_}{\textbf{Proposition}}
\newtheorem{corollary_}{\textbf{Corollary}}
\usepackage{xcolor}
\usepackage{algorithm}
\usepackage{algorithmic}
\usepackage{subcaption}
\usepackage[inline]{enumitem}
\DeclareMathOperator*{\argmin}{argmin}
\title{On transferring safety certificates across dynamical systems}
\author[1]{Nikolaos Bousias}
\author[1]{Charalampia Stamouli}
\author[2]{Anastasios Tsiamis}
\author[1]{George Pappas}
\affil[1]{GRASP Lab, University of Pennsylvania}
\affil[2]{Automatic Control Laboratory, ETH Zürich}

\date{}

\begin{document}
\maketitle

\begin{abstract}
Control barrier functions (CBFs) provide a powerful tool for enforcing safety constraints in control systems, but their direct application to complex, high-dimensional dynamics is often challenging. In many settings, safety certificates are more naturally designed for simplified or alternative system models that do not exactly match the dynamics of interest. This paper addresses the problem of transferring safety guarantees between dynamical systems with mismatched dynamics. We propose a transferred control barrier function (tCBF) framework that enables safety constraints defined on one system to be systematically enforced on another system using a simulation function and an explicit margin term. The resulting transferred barrier accounts for model mismatch and induces a safety condition that can be enforced on the target system via a quadratic-program-based safety filter. The proposed approach is general and does not require the two systems to share the same state dimension or dynamics. We demonstrate the effectiveness of the framework on a quadrotor navigation task with the transferred barrier ensuring collision avoidance for the target system, while remaining minimally invasive to a nominal controller. These results highlight the potential of transferred control barrier functions as a general mechanism for enforcing safety across heterogeneous dynamical systems.

\end{abstract}

\section{Introduction}

As autonomous systems become increasingly prevalent across robotics, aerospace, and industrial automation, ensuring their safe operation remains a fundamental challenge. Control Barrier Functions (CBFs) have emerged as a powerful mathematical framework for enforcing safety constraints in real-time control systems~\cite{ames2019control, ames2017control}. By encoding safety requirements as forward-invariant sets, CBFs enable the synthesis of minimally invasive safety filters that can be integrated with arbitrary nominal controllers while providing formal guarantees of collision avoidance and constraint satisfaction.

Despite their theoretical elegance and practical success, a significant gap exists between the abstract models used for safety analysis and the physical systems on which these guarantees must hold. Safety specifications are often most naturally expressed and verified on simplified, nominal models---such as single or double integrators---that capture essential system dynamics while remaining tractable for formal verification. However, deploying these safety guarantees on actual robotic platforms introduces substantial complexity: nonlinear dynamics, underactuation, geometric constraints, and actuator limitations all conspire to invalidate safety certificates derived from idealized models.

The classical approach to bridging this gap relies on approximate bisimulation and simulation relations, which have proven instrumental in hierarchical verification and abstraction-based control synthesis~\cite{girard2007approximation, tabuada2009verification, reissig2017feedback}. These relations formalize the notion that one system (the concrete, physical system) can approximately reproduce the behavior of another (the abstract, nominal system). While simulation relations have been successfully applied to verify temporal logic specifications~\cite{kloetzer2008fully} and synthesize correct-by-construction controllers~\cite{rungger2017computing}, their application to safety-critical control via CBFs remains relatively unexplored. Existing CBF synthesis methods typically require designing barrier functions directly on the full-order nonlinear system~\cite{xiao2019control, nguyen2016exponential}, which can be computationally prohibitive or require significant domain expertise.

Recent work has begun to address the transferability of safety specifications across system abstractions. The concept of backup barrier functions~\cite{ames2014control} and the composition of multiple CBFs~\cite{borrmann2015control} provide mechanisms for handling complex safety requirements, while learning-based approaches~\cite{dawson2023safe, robey2020learning,bousias2025deepequivariantmultiagentcontrol} attempt to discover CBF candidates for high-dimensional systems. However, these methods do not explicitly leverage the relationship between abstract and concrete system models, potentially discarding valuable structure that could simplify the synthesis problem.

\subsection*{Contributions}
In this paper, we develop a formal framework for transferring Control Barrier Functions from abstract systems to concrete systems linked by simulation relations. Our key contributions are summarized below:

\begin{enumerate}
    \item \textbf{Transfer construction:} We propose a transferred control barrier function (tCBF) formulation that enables safety certificates designed on an abstract dynamical model to be systematically enforced on a higher-dimensional nonlinear system using a simulation function and an explicit margin term.
    \item \textbf{Formal guarantees \& design guidelines:} We prove that with appropriate design of the margin function, the transferred safety certificates are valid for the new system, thereby inheriting safety guarantees. We provide explicit conditions for selecting the margin function, including closed-form solutions for common cases. These guidelines enable practical implementation without requiring exhaustive numerical optimization.
\end{enumerate}
We apply the proposed framework to a quadrotor navigating a cluttered three-dimensional environment using a double-integrator abstraction, demonstrating how safety can be transferred from a simple abstract model to full nonlinear quadrotor dynamics. Our framework addresses a critical gap in compositional safety verification: it allows practitioners to derive safety specifications on tractable abstract models while providing rigorous guarantees when deployed on complex physical systems. This is particularly relevant for modern robotic platforms where computing resources at deployment time are limited, making it desirable to perform complex safety synthesis offline on simplified models.

% The remainder of this paper is organized as follows. Section \ref{section:preliminaries} introduces the control-theoretic preliminaries, including Control Barrier Functions, simulation functions, and the witness map construction used throughout the paper. In Section \ref{section:tCBF}, we present the transferred barrier construction and establish sufficient conditions under which safety certificates can be propagated across dynamical systems. The main result shows how appropriate margin design compensates for simulation mismatch. Technical details regarding the existence and regularity of the witness map and margin function are deferred to the Appendix.

\section{Preliminaries}\label{section:preliminaries}
% \subsection{Notation}
Consider dynamical systems 
\begin{align}
    &\Sigma_1: \dot{x}_1(t)=f_1(x_1(t),u_1(t))\in \mathfrak{X}(\mathcal{X}_1 \times \mathcal{U}_1)\,,\,y_1(t)=g_1(x_1(t)) \nonumber\\
    &\Sigma_2: \dot{x}_2(t)=f_2(x_2(t),u_2(t))\in \mathfrak{X}(\mathcal{X}_2\times \mathcal{U}_2)\,,\,y_2(t)=g_2(x_2(t)) \nonumber
\end{align}
with states $x_1 \in \mathcal{X}_1, x_2 \in \mathcal{X}_2$, inputs $u_1\in \mathcal{U}_1, u_2\in \mathcal{U}_2$, outputs on the same space $y_1,y_2\in \mathcal{Y}$ and $\mathfrak{X}(\mathcal{X}_{i})$ denoting the generated vector fields.

For a continuously differentiable function \(b_{i}:\mathcal{X}_{1,2}\to\mathbb R\), its Lie derivative along \(f_{1,2}\) is denoted by
\[
\mathcal{L}_{u_{i} \mapsto f_{i}}b_{i}(x_{i}) := \langle \nabla b_{i}(x_{i})^\top, f_{i}(x_{i},u_{i}) \rangle
\]

A function \(\alpha:\mathbb R_{\ge 0}\to\mathbb R_{\ge 0}\) is class-\(\mathcal K\) if it is continuous, strictly increasing, and \(\alpha(0)=0\). It is \(\mathcal K_\infty\) if, in addition, \(\alpha(s)\to\infty\) as \(s\to\infty\).

\subsection{Control Barrier Functions}
\begin{definition_}\label{def:cbf}
    A continuously differentiable function \(b_{i}: \mathcal{X}_{i}\to\mathbb R\) is a valid Control Barrier Function (CBF) for \(\Sigma_{i}\) in the closed set \(S_{i}:=\{x_{i}\in \mathcal{X}_{i}: b_{i}(x_{i})\ge 0\}\), if there exists \(\alpha_b\in\mathcal K\) such that:
\begin{align}\label{eq:cbf_def}
    \sup_{u_i\in U_i} \mathcal{L}_{u_i \mapsto f_i}b_i(x_i) \ge -\alpha_b\big(b_i(x_i)\big)\qquad \forall x_i\in \mathcal{X}_i
\end{align}
Under standard regularity (e.g., locally Lipschitz-bounded vector fields), any policy $u_i\in \mathcal{U}_i$ that enforces $$\mathcal{L}_{u_i(t) \mapsto f_i}b_i(x_i(t)) \ge -\alpha_b\big(b_i(x_i(t))\big)\qquad \forall x_i\in \mathcal{X}_i,t\in \mathbb{R}_{+}$$ renders \(S_i\) forward invariant, i.e. keeps the system safe.
\end{definition_}

\subsection{Simulation functions and interfaces}
\begin{definition_}\label{def:simulation_function}
    A continuously differentiable function \(V:\mathcal{X}_1\times \mathcal{X}_2\to\mathbb R_{\ge 0}\) is a \emph{simulation function} from system \(\Sigma_2\) to system \(\Sigma_1\) if, for some \emph{control interface} $F: \mathcal{X}_1\times \mathcal{X}_2\times \mathcal{U}_2 \to \mathcal{U}_1$,  $\forall (x_1,x_2,u_2) \exists$:
\begin{itemize}
    \item a class-\(\mathcal K_\infty\) function \(\underline\gamma\) bounding an output mismatch
    \begin{align}
        ||y_1-y_2||_{\mathcal{Y}}\leq \underline\gamma(V(x_1,x_2)
    \end{align}
    \item a class-\(\mathcal K_\infty\) function \(\alpha_V\) such that the \emph{decay inequality} holds:
    \begin{align}
    \langle \nabla_{x_1}V(x_1,x_2)^\top, f_1(x_1,F(x_1,x_2,u_2)\rangle
    + \langle \nabla_{x_2}V(x_1,x_2)^\top, f_2(x_2,u_2)\rangle 
    \le -\alpha_V\big(V(x_1,x_2)\big)
    \end{align}\label{eq:simulation_decay}
\end{itemize}
\end{definition_}
For simplicity we drop the time dependence from the notation and leave it implied.
\subsection{Witness map and the pushforward on the witness graph}
A (local) \emph{witness map} \(\Pi: X_2\to X_1\) assigns an element from $\mathcal{X}_1$ to \(x_2\) via
\begin{align}\label{eq:witness_map}
    \Pi(x_2)\in\arg\min_{x_1} V(x_1,x_2)
\end{align}
We assume that \(\Pi\) is continuously differentiable on a neighborhood of interest and that the minimizer is unique there. Under these conditions, first-order optimality yields
\begin{align}\label{eq:optimality}
    \nabla_{x_1}V\big(\Pi(x_2),x_2\big) = 0
\end{align}

We define the \emph{witness graph} $\mathcal G := \{(x_1,x_2)\in \mathcal{X}_1\times \mathcal{X}_2: x_1=\Pi(x_2)\}$ with its tangent space at \((\Pi(x_2),x_2)\) denoted $T_{(\Pi(x_2),x_2)}\mathcal G = \{\,(D\Pi(x_2)v,\ v) : v\in T_{x_2}\mathcal{X}_2\,\}$.

The pair of vector fields \(\big(f_1(x_1,F(x_1,x_2,u_2)),\ f_2(x_2,u_2)\big)\) is \emph{tangent to} \(\mathcal G\) at \((\Pi(x_2),x_2)\) if
\begin{align}\label{eq:mismatch}
    D\Pi(x_2)\,\big[f_2(x_2,u_2)\big] 
= f_1\Big(\Pi(x_2),F\big(\Pi(x_2),x_2,u_2\big)\Big)
\end{align}
When (\ref{eq:mismatch}) holds, the chain rule collapses to a directional derivative in the \(\Sigma_1\) dynamics with the interface:
\begin{align}\label{eq:cbf_dynamics_maching}
    \langle \nabla b_1\big(\Pi(x_2)\big)^\top, D\Pi(x_2)\,f_2(x_2,u_2)\rangle
= \mathcal{L}_{F\big(\Pi(x_2),x_2,u_2\big) \mapsto f_1}b_1\big(\Pi(x_2)\big)
\end{align}
which is the \emph{pushforward identity on the witness graph}. The witness graph formalizes the notion that each concrete state is paired with its closest abstract representative under the simulation metric. When the joint dynamics are tangent to this graph, evolution of the concrete system can be interpreted through the abstract dynamics via a pushforward relationship. This observation allows Lie derivatives on the concrete system to be related directly to those on the abstract system, forming the backbone of the safety transfer analysis.

\section{Transferring safety certificates across dynamical systems}\label{section:tCBF}
The central challenge in transferring safety certificates is that the abstract and concrete systems generally evolve on different state spaces and under different dynamics. Even when their behaviors are related through a simulation function, enforcing a barrier constraint directly on the concrete system may fail due to mismatch between the two trajectories.

Our approach compensates for this mismatch by shrinking the abstract safe set according to the simulation error. Intuitively, if the concrete state is close (in the sense of the simulation function) to an abstract safe state, then safety can still be guaranteed, provided a suitable margin is enforced. This idea motivates the following construction.
\begin{definition_}
    Consider a valid CBF for system \(\Sigma_1\), $b_1:\mathcal{X}_1\rightarrow \mathbb{R}$ of Definition \ref{def:cbf}. For some simulation function $V:\mathcal{X}_1\times \mathcal{X}_2\to\mathbb R_{\ge 0}$ with control interface $F: \mathcal{X}_1\times \mathcal{X}_2\times \mathcal{U}_2 \to \mathcal{U}_1$ from Definition \ref{def:simulation_function}, we define the \emph{transferred Control Barrier Function} (t-CBF) $b_2:\mathcal{X}_2\rightarrow \mathbb{R}$ from $\Sigma_1$ on $\Sigma_2$:
\begin{align}\label{eq:t-CBF}
     b_2(x_2) := b_1\big(\Pi(x_2)\big) - \phi\big(V(\Pi(x_2),x_2)\big), \qquad \phi\in\mathcal K_\infty
\end{align}
\end{definition_}\label{def:t-cbf}

% \subsection{Lower bound on the transferred barrier derivative}
\begin{definition_}\label{def:dynamic_mismatch}
    For $V:\mathcal{X}_1\times \mathcal{X}_2\to\mathbb R_{\ge 0}$ and $F: \mathcal{X}_1\times \mathcal{X}_2\times \mathcal{U}_2 \to \mathcal{U}_1$ (Definition \ref{def:simulation_function}) and induced witness map $\Pi: X_2\to X_1$ of (\ref{eq:witness_map}), define the \emph{dynamics mismatch}:
    \begin{align}
        \delta(x_2,u_2) := D\Pi(x_2)\,[f_2(x_2,u_2)] - f_1\Big(\Pi(x_2)\,,F\big(\Pi(x_2),x_2,u_2 \big)\Big)
    \end{align}
\end{definition_}

\begin{assumption_}\label{ass:mismatch_bound}
One of the following hypotheses stands.
\begin{enumerate}
\item[H1.] Exact pushforward: If (\ref{eq:mismatch}) holds, then \(\delta\equiv 0\).
\item[H2.] Stationarity: If (\ref{eq:optimality}) holds \(\exists\gamma\in\mathcal K_\infty\) with $|\nabla_{\Pi} b_1(\Pi(x_2))^\top\!\delta(x_2,u_2)|\le \gamma\big(V(\Pi(x_2),x_2)\big)$
\item[H3.] General approximate pushforward: if \(\exists\Gamma\in\mathcal K_\infty\) with $$\Big|\Big(\nabla_\Pi b_1\big(\Pi(x_2)\big) - d\phi\Big(V(\Pi(x_2),x_2)\Big)\,\nabla_{x_1}V(\Pi(x_2),x_2)\Big)^\top \delta(x_2,u_2)\Big|\le \Gamma\big(V(\Pi(x_2),x_2)\big)$$
\end{enumerate}
\end{assumption_}

We now analyze how the transferred barrier evolves along trajectories of the concrete system. The key idea is to decompose the time derivative of the transferred barrier into the following components:
\begin{enumerate*}[label=\roman*)]
    \item a term corresponding to the abstract CBF dynamics
    \item a term inherited from the simulation function decay
    \item a residual mismatch term capturing imperfect pushforward
\end{enumerate*}
By bounding each component separately, we obtain a differential inequality that isolates the role of the margin function in restoring the standard CBF condition.

\begin{lemma_}\label{lemma:db2_bound}
    Suppose \(b_1\) is a valid CBF for \(\Sigma_1\) with \(\alpha_b\in\mathcal K\), and \(V\) is a simulation function with decay \(\alpha_V\in\mathcal K_\infty\) and interface \(F\), then :
    \begin{align*}
        \dot b_2(x_2,u_2) \;\ge\; -\alpha_b\big(b_1(\Pi(x_2))\big) + d\phi\big(V(\Pi(x_2),x_2)\big)\,\alpha_V\big(V(\Pi(x_2),x_2)\big) - r\big(V(\Pi(x_2),x_2)\big)
    \end{align*}
    where \(r\in \mathcal{K}_\infty\) under Assumption \ref{ass:mismatch_bound} and $\forall x_2,u_2$.
\end{lemma_} 
\begin{proof}
    Differentiating the t-CBF \(b_2\) along trajectories of \(\Sigma_2\) yields, by the chain rule,
\begin{align}\label{eq:time_derivative_t-cbf}
\dfrac{d}{dt} b_2(x_2) &= \langle\nabla b_1\big(\Pi(x_2)\big)^\top, D\Pi(x_2)\,\big[f_2(x_2,u_2)\big] \rangle \nonumber\\
&\quad - d\phi\big(V(\Pi(x_2),x_2)\big)\Big( \langle\nabla_{x_1}V\big(\Pi(x_2),x_2\big)^\top, D\Pi(x_2)\,[f_2(x_2,u_2)] \rangle \nonumber\\
&\quad+ \langle\nabla_{x_2}V\big(\Pi(x_2),x_2\big)^\top ,[f_2(x_2,u_2)] \rangle\Big)
\end{align}
Lemma 1 shows that the transferred barrier satisfies a relaxed CBF condition, where violation of the abstract barrier constraint is compensated by decay of the simulation function and the margin derivative. The remaining task is therefore to choose the margin function so that these additional terms dominate the abstract barrier growth, recovering a standard CBF inequality on the concrete system.

From the optimality condition of the witness map (\ref{eq:optimality}) $\nabla_{x_1}V\big(\Pi(x_2),x_2\big) = 0$, thus the second line in \ref{eq:time_derivative_t-cbf} is simplified. We retain it in the proof below to accommodate scenarios where only approximate optimality is available.
Adding/subtract \(f_1(\Pi,F)\) in \ref{eq:time_derivative_t-cbf} wherever \(D\Pi[f_2]\) appears:

\begin{align}\label{eq:db2_CBF_simulation_mismatch}
&\dfrac{d}{dt} b_2(x_2) = 
\underbrace{\Big\langle \nabla_{\Pi} b_1(\Pi(x_2))^\top, f_1\Big(\Pi(x_2)\,,F\big(\Pi(x_2),x_2,u_2 \big)\Big) \Big\rangle}_{\text{CBF term on }\Sigma_1} \nonumber\\
&-d\phi(V(\Pi(x_2),x_2))\,\underbrace{\Big[\langle \nabla_{x_1}V(\Pi(x_2),x_2)^\top\,,f_1\Big(\Pi(x_2)\,,F\big(\Pi(x_2),x_2,u_2 \big)\Big)\rangle + 
\langle \nabla_{x_2}V(\Pi(x_2),x_2)^\top\,,f_2(x_2,u_2) \rangle\Big]}_{\text{simulation term}} \nonumber\\
&\quad + \underbrace{\Big(\nabla_\Pi b_1\big(\Pi(x_2)\big) - d\phi\Big(V(\Pi(x_2),x_2)\Big)\,\nabla_{x_1}V(\Pi(x_2),x_2)\Big)^\top \delta(x_2,u_2)}_{\text{mismatch term}}
\end{align}

\noindent For the CBF term on \(\Sigma_1\), if \(b_1\) is a valid CBF for \(\Sigma_1\), then $\mathcal{L}_{u_1\mapsto f_1}b_1(x_1)\geq \alpha_b(b_1(x_1))\,,\,\forall x_1,u_1$.
Evaluating at \(x_1=\Pi(x_2)\) with \(u_1 = F(\Pi(x_2),x_2,u_2)\) yields:
$$\langle \nabla_{\Pi} b_1(\Pi(x_2))^\top, f_1\Big(\Pi(x_2)\,,F\big(\Pi(x_2),x_2,u_2 \big)\Big) \rangle \geq \alpha_b\Big(b_1\big(\Pi(x_2)\big)\Big)$$

\noindent For the simulation term, the decay (\ref{eq:simulation_decay}) evaluated at \((x_1,x_2)=(\Pi(x_2),x_2)\) yields:
\begin{align*}
    \langle \nabla_{x_1}V(\Pi(x_2),x_2)^\top\,,f_1\Big(\Pi(x_2)\,,F\big(\Pi(x_2),x_2,u_2 \big)\Big)\rangle + 
\langle \nabla_{x_2}V(\Pi(x_2),x_2)^\top\,,f_2(x_2,u_2) \rangle \leq -\alpha_V\Big(V\big(\Pi(x_2)\big)\Big)
\end{align*}
Combining (\ref{eq:db2_CBF_simulation_mismatch}) with CBF/Simulation bounds and Assumption \ref{ass:mismatch_bound} yields the lower bound of Lemma \ref{lemma:db2_bound}.
\end{proof}

% \subsection{CBF transfer condition}
\noindent We now isolate a design condition on \(\phi\) that converts Lemma \ref{lemma:db2_bound} into a standard CBF inequality.

\begin{assumption_}\label{ass:phi}
For $\alpha_V,\alpha_b,r:[0,\infty)\to[0,\infty)$ be (extended) $\mathcal K$-class functions, i.e., continuous, strictly increasing, and $\alpha_V(0)=\alpha_b(0)=r(0)=0$ \(\exists\phi\in\mathcal K_\infty\) such that $\phi'(s)\,\alpha_V(s) \;\ge\; \alpha_b\big(\phi(s)\big) + r(s)\,\,,\,\, \forall s\geq0$.
\end{assumption_}
 Intuitively, the margin function must grow fast enough so that the decay of the simulation function offsets both the abstract barrier contraction and any residual dynamics mismatch. This condition depends only on scalar comparison functions, making margin synthesis independent of the system dimension.

\begin{theorem_}\label{thm:comparison}
Fix $s_0>0$ and $\phi(s_0)=\eta\ge 0$. Assume that for each compact $K\subset[0,\infty)$ the map
$x\mapsto \alpha_b(x)$ is locally Lipschitz on $K$.
Let $y:[s_0,\infty)\to[0,\infty)$ be the unique Carath\'eodory solution of
\begin{equation}\label{eq:comparison-ode}
\dot y(s)=\lambda(s,y(s))=\frac{\alpha_b(y(s))+r(s)}{\alpha_V(s)},\qquad y(s_0)=\eta.
\end{equation}
If $\phi$ is absolutely continuous, satisfies the ODE inequality for a.e.\ $s\ge s_0$, and $\phi(s_0)\ge \eta$, then
\begin{equation}\label{eq:phi-ge-y}
\phi(s)\;\ge\;y(s),\qquad \forall s\ge s_0.
\end{equation}
In particular, among all $\phi\in \mathcal{K}_\infty$ satisfying the the ODE inequality, the solution $y(s)$ is pointwise minimal.
\end{theorem_}
\begin{proof}
    Appendix C.
\end{proof}
Since $s\mapsto r(s)/\alpha_V(s)$ is locally integrable on $[s_0,\infty)$, because $r,\alpha_V\in\mathcal{K}_\infty$, and if  $a_b\in \mathcal{K}_\infty$ is locally Lipschitz (common), then by Picard–Lindelöf lemma (applies the Banach fixed-point theorem to Picard iterations) for the ODE \ref{eq:comparison-ode} there always exists a unique solution in $[s_0,s_0+\epsilon],s_0,\epsilon\in \mathbb{R}^{+}$, recovered by the Picard iterations:
\begin{align}
    y_{\kappa+1}(s)=\eta+\int_{s_0}^{s} \dfrac{a_b(y_\kappa(\tau))+r(\tau)}{a_V(\tau)} d\tau \qquad s\geq s_0,y_0=\eta
\end{align}

In the case of perfect dynamics recapturing, i.e. $r\equiv 0$, Assumption \ref{ass:phi} can be interpreted as a separation condition between safety enforcement and model mismatch. For some reference points $\hat{x},\hat{s}\in \mathbb{R}^{+}$, defining $F(x):=\int_{\hat{x}}^{x} \frac{1}{a_b(u)}du$ and $G(x):=\int_{\hat{s}}^{s} \frac{1}{a_V(\tau)}d\tau$, yields separated solution $y(s) = F^{-1}(F(\eta)+G(s)-G(s_0))$. Since extended 
$\mathcal{K}$ functions are strictly increasing, $\alpha_b$ is injective, and $F,G$ are strictly increasing wherever the integrals are well-defined, so $F^{-1}$ is well-defined on its image.

\begin{proposition_}\label{prop:lineardecay}
Let $\alpha_b(s) = c_b s$ and $\alpha_V(s) = c_V s$ with $c_b, c_V > 0$. 
Fix $s_0 > 0$ and $\phi(s_0) = \eta \ge 0$. 
Then the Carath\'eodory solution of the comparison ODE \ref{eq:comparison-ode} admits the representation
\begin{equation}
y(s)
=
\left(\frac{s}{s_0}\right)^{\frac{c_b}{c_V}} \eta
+
\frac{1}{c_V}\,
s^{\frac{c_b}{c_V}}
\int_{s_0}^{s}
\tau^{-\frac{c_b}{c_V}-1}\, r(\tau)\, d\tau,
\qquad s \ge s_0.
\end{equation}
In particular, if $r \equiv 0$, then
\begin{equation}
y(s) = \eta \left(\frac{s}{s_0}\right)^{\frac{c_b}{c_V}} .
\end{equation}
If  $r$ is linear, i.e. $r(s)=c_r s$ with $c_r\ge 0$, then
\begin{equation}
y(s)
=
\begin{cases}
\displaystyle \eta\left(\frac{s}{s_0}\right)^{\lambda}
+
\frac{c_r}{c_V(1-\lambda)}\left(s^{1-\lambda}-s_0^{1-\lambda}\right), & \lambda\neq 1,\\[8pt]
\displaystyle \eta\left(\frac{s}{s_0}\right) + \frac{c_r}{c_V}s\log\!\left(\frac{s}{s_0}\right), & \lambda = 1.
\end{cases}
\end{equation}
\end{proposition_}
\begin{proof}
    Appendix D.
\end{proof}
% For the simple linear-gain case where \(\alpha_b(s)=c_b s\) and \(\alpha_V(s)=c_V s\) with constants \(c_V\geq c_b>0\), and if \(r\equiv 0\), then any linear \(\phi(s)=c_\phi s\) with \(c_\phi > 0\) satisfies Assumption \ref{ass:phi}. If a residual bound \(r(s)\le c_r s\) is present, then one may choose \(c_\phi \ge c_r/(c_V-c_b)\).

\begin{theorem_}\label{theorem:validtcbf}
    Let Assumptions \ref{ass:mismatch_bound} and \ref{ass:phi} stand. There exists a class-\(\mathcal K\) function \(\widetilde\alpha:\mathbb R \rightarrow \mathbb R\) such that along trajectories of \(\Sigma_2\) it stands that $ \frac{d}{dt}b_2(x_2,u_2) + \widetilde\alpha\big(b_2(x_2)\big) \ge 0\,\,,\,\, \forall x_2 \in \mathcal{X}_2,u_2\in \mathcal{U_2}$, thus, any feedback enforcing this inequality renders \(S_2:=\{x_2| b_2(x_2)\ge 0\}\) forward invariant, i.e., \(b_2\) is a valid CBF for \(\Sigma_2\).
\end{theorem_}
\begin{proof}
    Using \(b_1\big(\Pi(x_2)\big) = b_2(x_2)+ \phi\big(V(\Pi(x_2),x_2)\big)\) in (\ref{eq:t-CBF}) and Assumption \ref{ass:phi},
\begin{align*}
    \dot b_2(x_2) \;\ge\; -\alpha_b\big(b_2(x_2)+\phi(V(\Pi(x_2),x_2))\big) + \underbrace{\big(d\phi(V(\Pi(x_2),x_2))\alpha_V(V(\Pi(x_2),x_2)) - r(V(\Pi(x_2),x_2))\big)}_{\ge\ \alpha_b(\phi(V(\Pi(x_2),x_2)))}\ 
\end{align*}
Define the comparison function $\widetilde\alpha(s) := \sup_{r\ge 0}\big(\alpha_b(s+r)-\alpha_b(r)\big)\,,\, s\ge 0$. 
Since \(\alpha_b\in\mathcal K\): (i) \(\widetilde\alpha(0)=0\); (ii) For \(s_2>s_1\), \(\alpha_b(s_2+r)-\alpha_b(r) > \alpha_b(s_1+r)-\alpha_b(r)\) for all \(r\), so after taking the supremum, \(\widetilde\alpha(s_2)>\widetilde\alpha(s_1)\); (iii) Continuity follows from monotonicity and continuity of \(\alpha_b\in \mathcal{K}_\infty\). Hence \(\widetilde\alpha\in\mathcal K\) and \(\alpha_b(s+r)-\alpha_b(r)\le \widetilde\alpha(s)\) for all \(r,s\ge 0\). Therefore, $\dot b_2 \;\ge\; -\widetilde\alpha\big(b_2\big)$. Forward invariance of \(S_2\) then follows from standard CBF arguments. 
\end{proof}
\begin{remark_}
    For the case exact push-forward, i.e. no dynamics miss-match $\delta=0$, System 1 constitutes a reduced-order model of System 2 and the t-CBF is similar to \cite{9652122,molnar2023safety}.
\end{remark_}
Theorem \ref{theorem:validtcbf} establishes that safety verified on an abstract model can be enforced on a more complex concrete system without redesigning the barrier function. The only additional requirement is the construction of a margin that quantifies simulation error. As a result, safety synthesis and safety enforcement can be cleanly decoupled across system representations.

\subsection*{Safe optimal control with tCBF-QP}
This section describes how the transferred Control Barrier Function (tCBF) is enforced in real time via a quadratic program (QP) on the target system $\Sigma_2$. Substituting Lemma~\ref{lemma:db2_bound} into Definition~\ref{eq:t-CBF} yields the constraint
\begin{align}
\left\langle \nabla b_1(\Pi(x_2)), D\Pi(x_2) f_2(x_2,u_2) \right\rangle
&-
\phi'(V)\,
\left\langle \nabla_{x_2} V(\Pi(x_2),x_2), f_2(x_2,u_2) \right\rangle
\nonumber\\
&\ge
-\alpha_e\!\left(b_2(x_2)\right)
+
r\!\left(V(\Pi(x_2),x_2)\right).
\label{eq:tcbf_qp_constraint}
\end{align}
For control-affine target dynamics $f_2(x_2,u_2) = f_{2,0}(x_2) + f_{2,1}(x_2) u_2$, the constraint~\eqref{eq:tcbf_qp_constraint} is action-affine and thus suitable for quadratic programming. Let $u_{\mathrm{nom}}(x_2)$ denote an arbitrary nominal controller.
The tCBF-based safety filter is given by the quadratic program
\begin{equation}
\begin{aligned}
u_2^\star(x_2) =
\arg\min_{u_2 \in U_2} \quad &
\|u_2 - u_{\mathrm{nom}}(x_2)\|^2 \\
\text{s.t.} \quad &
A(x_2)\,u_2 \;\ge\; b(x_2)
\end{aligned}
\end{equation}
where, for $\bar x_1 \triangleq \Pi(x_2)$, $\bar V \triangleq V(\bar x_1,x_2)$ and $\phi' \triangleq d\phi(\bar V)$,
\begin{align*}
A(x_2)&:=
\Big(\nabla b_1(\bar x_1)\Big)^\top D\Pi(x_2)\, f_{2,1}(x_2)
-
\phi'( V(\bar x_1,x_2))\, \nabla_{x_2} V(\bar x_1,x_2)^\top f_{2,1}(x_2) \\[4pt]
% \label{eq:tcbf_A}
b(x_2)
&:=
-\alpha_e\!\big(b_2(x_2)\big)
+
r(\bar V)
-
\Big(\nabla b_1(\bar x_1)\Big)^\top D\Pi(x_2)\, f_{2,0}(x_2)
+
\phi'( V(\bar x_1,x_2))\nabla_{x_2} V(\bar x_1,x_2)^\top f_{2,0}(x_2),
% \label{eq:tcbf_b}
\end{align*}
Any solution $u_2^\star$ renders the set $S_2 = \{ x_2 \in X_2 \mid b_1\big(\Pi(x_2)\big) - \phi\big(V(\Pi(x_2),x_2)\big) \ge 0 \}$ forward invariant.
\begin{algorithm}[h!]
\caption{tCBF-QP Safety Filter}
\label{alg:tcbf_qp}
\begin{algorithmic}[1]
\REQUIRE Current state $x_2$, nominal input $u_{\mathrm{nom}}(x_2)$
\ENSURE Safe control input $u_2^\star$
\vspace{0.5em}
\STATE Compute witness state $\Pi(x_2) \in \arg\min_{x_1} V(x_1,x_2)$
\STATE Evaluate $V \gets V(\Pi(x_2),x_2)$
\STATE Compute margin terms $\phi(V)$ and $\phi'(V)$
\STATE Evaluate gradients $\nabla b_1(\Pi(x_2))$, $D\Pi(x_2)$, and $\nabla_{x_2}V(\Pi(x_2),x_2)$
\STATE Form the affine constraint~\eqref{eq:tcbf_qp_constraint}
\STATE Solve the quadratic program: $\min_{u_2 \in U_2} \ \|u_2 - u_{\mathrm{nom}}\|^2
\quad
\text{s.t. } \dot b_2 + \alpha_e(b_2) \ge 0$
\STATE Apply $u_2^\star$
\end{algorithmic}
\end{algorithm}

\section{Experiments}

We consider the problem of safely navigating a quadrotor through a static, obstacle-cluttered environment from a given initial position $p_0\in\mathbb{R}^3$ to a desired target $p_T\in\mathbb{R}^3$ while avoiding collisions. Obstacles $\mathcal{O}=\{1,..,N\}$ are modeled as spheres with centers $c_i \in \mathbb{R}^3$ and inflated radii $\rho_i>0$. Let $\Sigma_1$ be the 3D double integrator on $\mathcal{X}_1=\mathbb{R}^3\times\mathbb{R}^3$ with dynamics $\dot p_1 = v_1\,,\,\dot v_1 = u_1, \,,\, y_1 = p_1$, where $x_1=(p_1,v_1)$ and $u_2\in\mathbb{R}^3$ is the abstract acceleration input.
Let $\Sigma_2$ be the standard rigid-body quadrotor on 
$\mathcal{X}_2=\mathbb{R}^3\times\mathbb{R}^3\times SO(3)\times\mathbb{R}^3$ with dynamics:
\begin{align*}
    \dot x_2= \dfrac{d}{dt}\begin{bmatrix} p_2\\v_2\\R\\\Omega\end{bmatrix} = \begin{bmatrix}
        v_2\\-g\,e_3 + \frac{1}{m}f\,R e_3\\ R\, [\Omega]_\wedge \\ J^{-1}\big(M-\Omega\times (J\Omega)\big)
    \end{bmatrix}
\end{align*}
and $y_2=p_2$, $m>0$, $g>0$, $e_3=[0\;0\;1]^\top$,  inputs $u_2=(f,M)\in \mathbb{R} \times \mathbb{R}^3$, and $ [\cdot]_\wedge: \mathbb{R}^3 \rightarrow \mathfrak{so(3)}$ is the hat operator.

\subsection{Policy interface, simulation function \& witness map}
Define the error states $e_p=p_2-p_1\,,\,e_v=v_2-v_1$ and consider the control interface $F(x_1,x_2,u_2)=(\frac{f}{m}Re_3-ge_3)+K_p\,e_p+K_v\,e_v$ for $K_p,K_v\in \mathbb{R}^{3\times3}$ such that $\mathbb{A}=\begin{bmatrix}
0&I\\
-K_p&-K_v
\end{bmatrix}$ is Hurwitz stable. For some $\mathbb{Q}\succ0$ let $\mathbb{P}\succ0$ be the unique solution of the Lyapunov equation $\mathbb{A}^T\mathbb{P}+\mathbb{P}\mathbb{A}=\mathbb{Q}$. Define the continuously differentiable positive semi-definite candidate simulation function $V(x_1,x_2):= z^\top P z \geq 0\,,\, z=[e_p,e_v]^T\in\mathbb{R}^6$ with witness map $\Pi(x_2)=\argmin_{x_1}V(x_1,x_2)=[p_2,v_2]^T$ satisfying the first-order optimality condition $\nabla_{x_1}V(\Pi(x_2),x_2)=0$. The candidate simulation function is valid according to Definition \ref{def:simulation_function} since $\lVert y_2-y_1 \rVert=\lVert e_p \rVert \leq \gamma(V(x_1,x_2))$ for $\gamma(s)=\dfrac{1}{\sqrt{\lambda_{\min}(\mathbb{P})}}\sqrt{s} \in \mathcal{K}_\infty$ and from the error dynamics the decay inequality is $\dot{V}(x_1,x_2)=-z^T\mathbb{Q}z\leq -\lambda_{\min}(\mathbb{Q})\lVert z\rVert^2\leq -\alpha_V(V(x_1,x_2))$ for $\alpha_V(s)=\dfrac{\lambda_{\min}(\mathbb{Q})}{\lambda_{\max}(\mathbb{P})}s\in \mathcal{K}_\infty$. The dynamics mismatch, then, is $\delta(x_2,u_2) := D\Pi(x_2)\,[f_2(x_2,u_2)] - f_1\Big(\Pi(x_2)\,,F\big(\Pi(x_2),x_2,u_2 \big)\Big)\equiv 0$ satisfying Assumption \ref{ass:mismatch_bound}.

\subsection{Transfering safety certificates from $\Sigma_1$ to $\Sigma_2$}
Define the distance-squared safety function $h_i(x_1) := \|p_1 - c_i\|^2 - \rho_i^2\,,\,\forall i\in \mathcal{O}$. For $\Sigma_1$ the CBF candidate $h_i$ has relative degree 2 and safety is enforced via an exponential CBF condition $\ddot h_i(x_1) + k_{b,1} \dot h_i(x_1) + k_{b,2} h_i(x_1) \ge 0$ which is equivalent to enforcing a zeroing CBF $b_{1,i}(x_1) := \dot h_i(x_1) + k_{b,1} h_i(x_1)$ satisfying $\dot b_{1,i}(x_1,u_1) \ge -\alpha_b\!\left(b_{1,i}(x_1)\right)$ for $\alpha_b(s) := k_{b,2} s \in \mathcal{K}_\infty,k_{b,2},k_{b,2}>0$, yielding an affine constraint on the abstract control input $u_1$. To deploy this safety certificate on a quadrotor with nonlinear, underactuated dynamics, we construct the candidate tCBF from \ref{eq:t-CBF}. With an appropriate choice of $\phi$ satisfying the ODE inequality of Assumption \ref{ass:phi}, the function $b_{2,i}$ is a valid control barrier function for the quadrotor, enabling safe navigation via real-time quadratic programming according to Algorithm \ref{alg:tcbf_qp}. From Proposition \ref{prop:lineardecay}, since $\alpha_V,\alpha_b\in\mathcal{K}_\infty$ are linear and $\delta\equiv 0$, $\phi(s) \geq \phi(s_0) \left(\dfrac{s}{s_0}\right)^{\frac{k_{b,2}\lambda_{\max}(\mathbb{P})}{\lambda_{\min}(\mathbb{Q})}}$ for any $\phi(s_0),s_0\in \mathbb{R}^{+}$.

\begin{figure}
     \centering
     % \begin{subfigure}[b]{0.5\textwidth}
     %     \centering
     %     \includegraphics[width=\textwidth]{figures/frame_1_1.png}
     % \end{subfigure}
     % \hfill
     \begin{subfigure}[b]{0.49\textwidth}
         \centering
         \includegraphics[width=\textwidth]{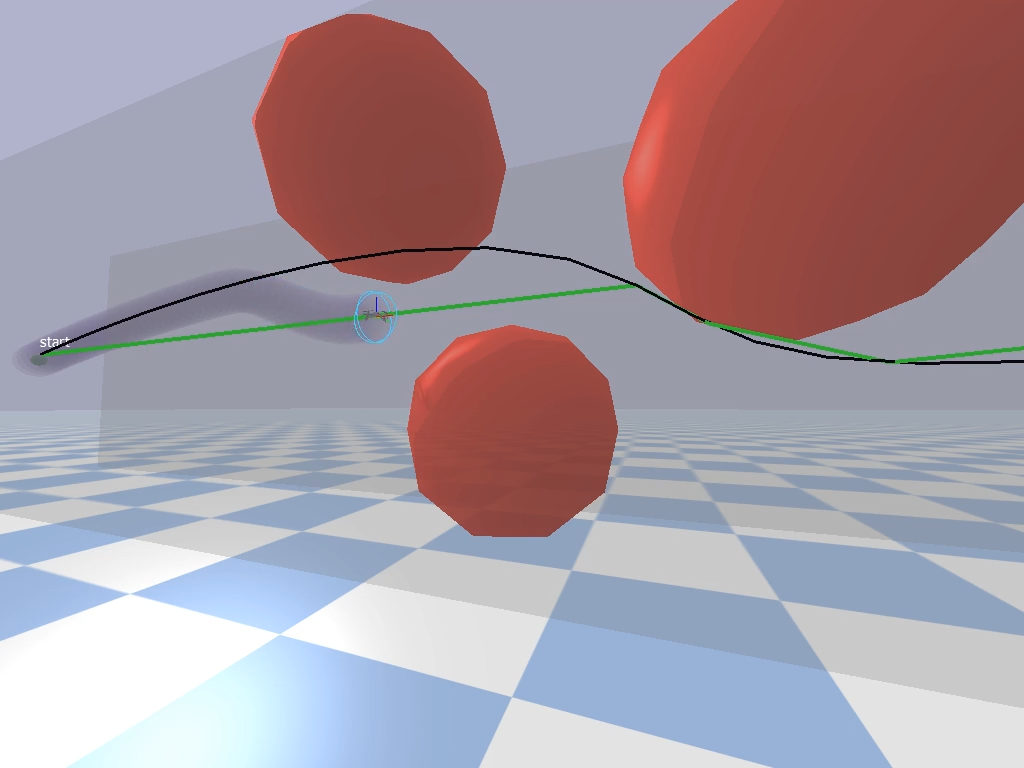}
     \end{subfigure}
     % \hfill
     \begin{subfigure}[b]{0.49\textwidth}
         \centering
         \includegraphics[width=\textwidth]{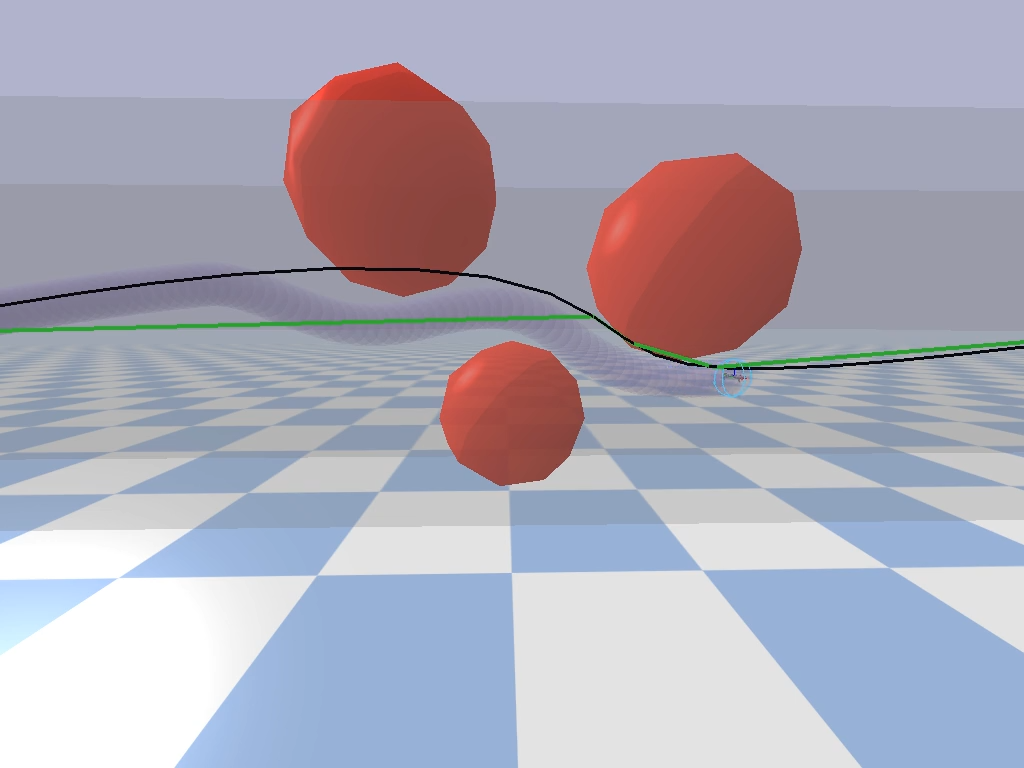}
     \end{subfigure}
        \caption{Instances of the quadrotor trajectory in an obstacle environment. Red spheres denote obstacles, the green curve shows the RRT* path, and the black curve shows the nominal minimum-snap trajectory incurring collisions. Semi-transparent purple spheres indicate the tCBF-QP trajectory.}
        \label{fig:3dtrajectories}
\end{figure}

\begin{figure}
     \centering
     \begin{subfigure}[b]{0.49\textwidth}
         \centering
         \includegraphics[width=\textwidth]{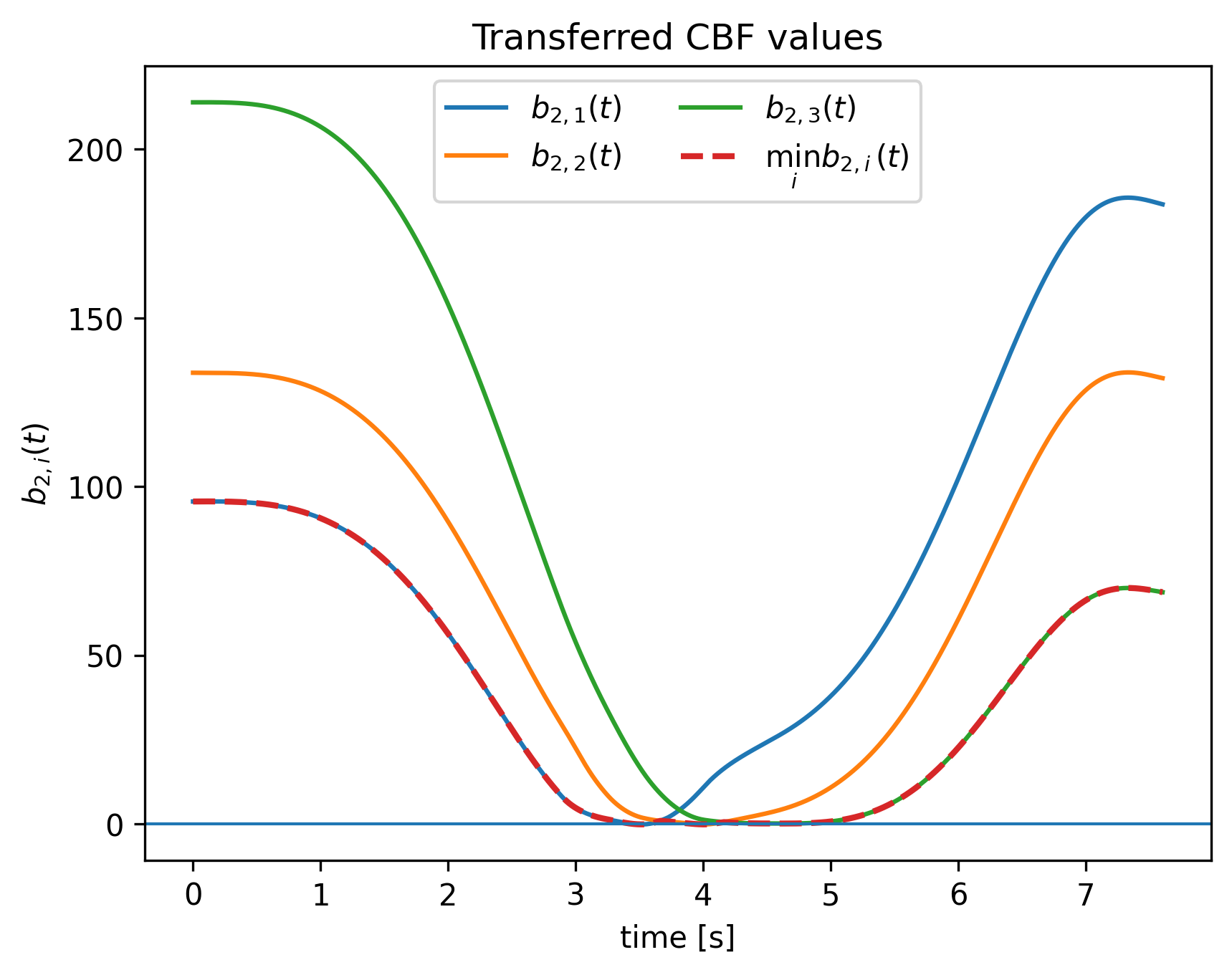}
         \label{fig:b2}
     \end{subfigure}
     \begin{subfigure}[b]{0.49\textwidth}
         \centering
         \includegraphics[width=\textwidth]{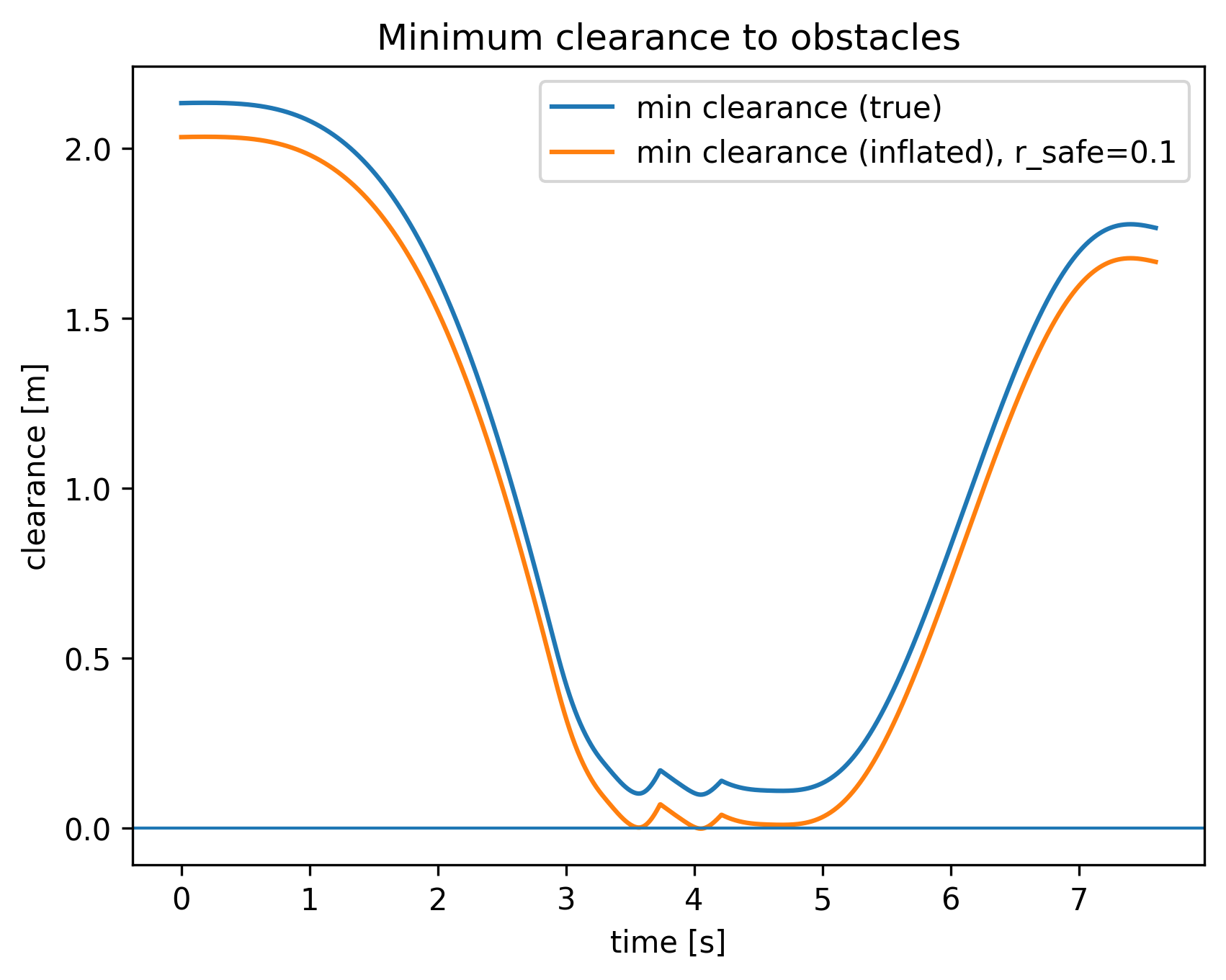}
         \label{fig:clearence}
     \end{subfigure}
        \caption{(Left) Evolution of the transferred control barrier functions associated with each obstacle. The minimum transferred barrier value is shown by the dashed curve. As the quadrotor approaches obstacles, the transferred barriers decrease and become active near zero, demonstrating enforcement of safety at the concrete system level despite abstraction mismatch.
        (Right) Minimum obstacle clearance along the trajectory. The true geometric clearance remains strictly positive, confirming collision avoidance. The inflated clearance used in the abstract barrier construction is shown for reference, illustrating the additional safety margin imposed by the transferred CBF.}
        \label{fig:b2-clearence}
\end{figure}
\subsection{Nominal quadrotor policy design}
The transferred control barrier function is implemented as a safety filter that minimally modifies a nominal quadrotor controller. The nominal policy is designed to generate dynamically feasible trajectories that drive the vehicle from a given start position to a target while respecting quadrotor dynamics in the absence of safety constraints. Our nominal control pipeline consists of motion planning, trajectory optimization, and geometric tracking control.
Initially, a sampling-based planner using RRT$^\ast$ \cite{doi:10.1177/0278364911406761} computes a collision-free geometric path from the start to the goal in the obstacle-cluttered environment. The resulting sequence of waypoints provides a feasible but non-smooth reference path. The discrete RRT$^\ast$ path is converted into a smooth, time-parameterized trajectory by solving a minimum-snap trajectory optimization problem \cite{Richter2016}. Specifically, utilizing the differential flatness of the quadrotor, we generate a piecewise polynomial trajectory $p_d(t)$ by minimizing the snap over the trajectory
\[
\min_{p_d(t)} \int_0^T \left\| \frac{d}{dt^4}p_d(t) \right\|^2 dt
\]
subject to continuity constraints on position and its derivatives up to snap, waypoint interpolation constraints, and boundary conditions on position, velocity, and acceleration for desired yaw aligned with the quadrotor's velocity, i.e. $\psi_d(t) = \text{acrtan}(v_{2,y},v_{2,x})$. This yields a smooth reference trajectory $p_d(t)$ with associated desired velocity $\dot{p}_d(t)$, acceleration $\ddot{p}_d(t)$, and jerk $p^{(3)}_d(t)$. Finally, the quadrotor tracks the reference trajectory using an $SE(3)$-geometric controller \cite{5980409}. The desired total force vector is computed as $F_d(t) = m\,( \ddot{p}_d(t) +g e_3)- \bar{k}_p (p_2(t) - p_d(t)) - \bar{k}_v (v(t) - \dot{p}_d(t)) $, for some $\bar{k}_p,\bar{k}_v>0$. The commanded thrust is $f_{\mathrm{nom}}(t) = F_d(t)^\top R(t) e_3\in \mathbb{R}$ and the desired body $z$-axis is defined as $b_{z,d}(t) = \frac{F_d(t)}{\|F_d(t)\|}$.
Together with the desired yaw direction, this defines a desired rotation matrix $R_d(t) = [b_{x,d},b_{y,d},b_{z,d}] \in SO(3)$ for $\sigma(t) = [\cos(\psi_d(t)),\sin(\psi_d(t)),0]^\top\,,\,b_{y,d}(t) = \frac{ b_{z,d}(t) \times \sigma(t)}{\lVert b_{z,d}(t) \times \sigma(t) \rVert}\,,\, b_{x,d}(t) = b_{y,d}(t) \times b_{z,d}(t)$. 
Let $e_R(t) = \frac{1}{2}[R_{d}^\top(t)R(t)-R^\top(t)R_{d}(t)]_\vee$ and $e_\Omega(t) = \Omega(t) - \dot{R}_d(t)$ denote the attitude and angular velocity error states, respectively, with $\bar{k}_R,\bar{k}_\Omega>0$ control gains and $ [\cdot]_\vee: \mathfrak{so(3)} \rightarrow \mathbb{R}^3$ denoting the vee operator.
The attitude control moment is then $M_{\mathrm{nom}}(t) = J(-\bar{k}_R e_R(t) - \bar{k}_\Omega e_\Omega(t))\in \mathbb{R}^3$, resulting to nominal input $u_{\mathrm{nom}}(t) = (f_{\mathrm{nom}}(t), M_{\mathrm{nom}}(t))$.
This nominal controller achieves aggressive trajectory tracking in free space, while the tCBF-based quadratic program modifies $u_{\mathrm{nom}}(t)$ only when required to ensure obstacle avoidance. The safety-filtered thrust-moment actions are converted to rotor actuation $$\omega_\mathrm{nom}^2(t)=\begin{bmatrix}
    k_F &k_F &k_F &k_F \\
    0 & k_Fl & 0 & -k_Fl \\
    -k_Fl & 0 & k_Fl & 0 \\
    -k_M & k_M & -k_M & k_M
\end{bmatrix}^{-1}\begin{bmatrix}
    f_\mathrm{nom}(t)\\ M_\mathrm{nom}(t)
\end{bmatrix}\in\mathbb{R}^4$$

\subsection{Results \& Discussion}
We now discuss the behavior of the proposed tCBF-QP safety filter and evaluate its ability to enforce safety on the full quadrotor dynamics while minimally modifying the nominal controller.

\noindent\textbf{Trajectory-level behavior :} Figure \ref{fig:3dtrajectories} shows representative snapshots of the quadrotor trajectory in the obstacle environment. The green curve corresponds to the collision-free geometric path computed by RRT*, while the black curve denotes the nominal minimum-snap trajectory obtained from trajectory optimization and tracked by the SE(3) controller in the absence of safety constraints. As expected, the nominal trajectory intersects the obstacle regions and would result in collisions. When the transferred control barrier function is enforced via the quadratic program, the resulting trajectory (visualized by the semi-transparent purple spheres) deviates locally from the nominal trajectory in the vicinity of obstacles while remaining close to it elsewhere. This demonstrates that the tCBF-QP acts as a minimally invasive safety filter: control modifications occur only when the system approaches the boundary of the transferred safe set, and nominal performance is largely preserved away from safety-critical regions. The quadrotor is able to pass through narrow regions between obstacles while remaining within the transferred safety envelope, highlighting the non-conservative nature of the proposed approach.

\noindent\textbf{Transferred barrier activation :} The evolution of the transferred control barrier functions associated with each obstacle is shown in Figure \ref{fig:b2-clearence} (left). Each curve corresponds to a transferred barrier, while the dashed curve represents their pointwise minimum. As the quadrotor approaches the obstacles, the corresponding transferred barrier values decrease and become active near zero, indicating enforcement of the safety constraint at the concrete system level. After the point of closest approach, the barrier values increase again as the system moves away from the obstacles. This behavior is consistent with the theoretical analysis in Section 3: although the safety certificate is synthesized on the abstract double-integrator model, the transferred barrier correctly identifies safety-critical situations for the full quadrotor dynamics and activates only when required.

\noindent\textbf{Collision avoidance :} Figure \ref{fig:b2-clearence} (right) reports the minimum distance between the quadrotor and the nearest obstacle over time. The true geometric clearance remains strictly positive throughout the trajectory, confirming collision avoidance for the full nonlinear system. For comparison, the inflated clearance used in the abstract barrier construction is also shown. Notably, the inflated clearance briefly approaches zero near the point of closest approach, reflecting the conservatism introduced to account for abstraction mismatch. Nevertheless, the true clearance remains positive, indicating that the transferred barrier successfully compensates for model mismatch while preserving physical safety. This observation empirically validates the role of the margin function $\phi$ in shrinking the abstract safe set to obtain a sound safety certificate on the concrete system.

Taken together, the results demonstrate that the proposed tCBF framework enables safety certificates synthesized on a simplified abstract model to be enforced on a high-dimensional, underactuated quadrotor system. The transferred barrier activates precisely in safety-critical situations, ensures collision avoidance in the concrete system, and results in only localized deviations from the nominal trajectory. These findings support the theoretical guarantees established in Section 3 and illustrate the practical effectiveness of transferred control barrier functions for safe autonomous navigation under model mismatch.

\section{Conclusions \& Future work}
This paper introduced a transferred control barrier function (tCBF) framework for enforcing safety guarantees using safety certificates synthesized on simpler abstract models. By leveraging a simulation function and an appropriate margin term, safety constraints defined on an abstract system can be systematically transferred to new  dynamical systems and enforced through a QP-based safety filter. Numerical experiments on a quadrotor navigating a cluttered three-dimensional environment demonstrated that the transferred barrier activates precisely in safety-critical situations, ensures collision avoidance for the full nonlinear system, and remains minimally invasive away from obstacles.

A key challenge in applying transferred barrier functions in practice is the construction of suitable simulation functions and policy interfaces, which can be difficult to derive analytically for complex systems. An important direction for future work is therefore to make these components learnable from data. Learning simulation functions and abstraction interfaces could significantly broaden the applicability of tCBFs and enable safety transfer in settings where accurate analytical models are unavailable.

\newpage
\appendix
\section*{Appendix}
The following appendices collect standard technical results that support the main construction and are included for completeness.
\subsection*{A: Constructing \texorpdfstring{\(\Pi\)}{Pi} and verifying (\ref{eq:optimality})}
Suppose \(V(\cdot,x_2)\) is strictly convex and \(C^2\) on \(X_1\) for each fixed \(x_2\) and coercive. Then the argmin is unique, and the implicit function theorem applied to \(\nabla_{x_1}V(x_1,x_2)=0\) around any point with \(\det(\nabla^2_{x_1}V)>0\) yields a \(C^1\) mapping \(\Pi(x_2)\). Equation (4) is precisely the first-order optimality condition.

\subsection*{B: On the existence of \texorpdfstring{\(\phi\)}{phi} satisfying Assumption \ref{ass:phi}}
Under local Lipschitzness of the right-hand side of the ODE and positivity of \(\alpha_V\) on \((0,\infty)\), the ODE has a unique maximal solution with \(\phi\in\mathcal K_\infty\) (strict increase follows from positivity; unboundedness follows from divergence of the integral \(\int^\infty \frac{\mathrm d s}{\alpha_V(s)}\) or by continuation of solutions). Any such solution satisfies the ordinary differential inequality as an equality; replacing equality with inequality preserves the result by monotonicity.

\subsection*{C: Proof of Theorem \ref{thm:comparison}}
\begin{proof}
Rewrite the ODE inequality on $[s_0,\infty)$ as $\dot\phi(s)\;\ge\;f(s,\phi(s))\; \text{for a.e.\ } s\ge s_0$, with $f$ the righ-hand side of the ODE inequality. Let $y$ solve \eqref{eq:comparison-ode}. Define $w(s):=(y(s)-\phi(s))_+=\max\{y(s)-\phi(s),0\}$.
For a.e.\ $s$ such that $w(s)>0$ we have $y(s)>\phi(s)$, hence by monotonicity of $\alpha_b$,
\[
\alpha_b(y(s))-\alpha_b(\phi(s))>0,
\]
and therefore
\[
\dot y(s)-\dot\phi(s)
\;=\;\frac{\alpha_b(y(s))+r(s)}{\alpha_V(s)}-\dot\phi(s)
\;\le\;\frac{\alpha_b(y(s))+r(s)}{\alpha_V(s)}-\frac{\alpha_b(\phi(s))+r(s)}{\alpha_V(s)}
\;=\;\frac{\alpha_b(y(s))-\alpha_b(\phi(s))}{\alpha_V(s)}.
\]
Using local Lipschitzness of $\alpha_b$ on bounded sets, there exists $L>0$ (on any interval where $y,\phi$ remain bounded) such that
$|\alpha_b(y)-\alpha_b(\phi)|\le L|y-\phi|=Lw$. Thus, for a.e.\ $s$,
\[
\frac{d}{ds}w(s)\;\le\;\frac{L}{\alpha_V(s)}\,w(s).
\]
Since $\alpha_V(s)>0$ for $s\ge s_0>0$ and $1/\alpha_V$ is locally integrable on $[s_0,\infty)$, Gr\"onwall's inequality yields
\[
w(s)\;\le\;w(s_0)\exp\!\left(\int_{s_0}^{s}\frac{L}{\alpha_V(\sigma)}\,d\sigma\right).
\]
But $w(s_0)=(y(s_0)-\phi(s_0))_+=(\eta-\phi(s_0))_+=0$ by $\phi(s_0)\ge\eta$, hence $w(s)\equiv 0$ and therefore $y(s)\le \phi(s)$ for all $s\ge s_0$.
\end{proof}

\subsection*{D: Proof of Proposition \ref{prop:lineardecay}}
\begin{proof}
With $\alpha_b(s)=c_b s$ and $\alpha_V(s)=c_V s$, the ODE on $[s_0,\infty)$ becomes, for $s>0$,
\begin{equation}
c_V s\,\phi'(s) = c_b \phi(s) + r(s),
\end{equation}
or equivalently
\begin{equation}
\phi'(s) - \frac{c_b}{c_V}\frac{1}{s}\phi(s) = \frac{1}{c_V}\frac{r(s)}{s}.
\end{equation}
Let $\lambda := c_b/c_V$ and consider the integrating factor $\mu(s)=s^{-\lambda}$, which is $C^1$ on
$[s_0,\infty)$ since $s_0>0$. Multiplying both sides by $\mu(s)$ yields
\begin{equation}
s^{-\lambda}\phi'(s) - \lambda s^{-\lambda-1}\phi(s) = \frac{1}{c_V}s^{-\lambda-1}r(s),
\end{equation}
and the left-hand side is the derivative of $s^{-\lambda}\phi(s)$, i.e.
\begin{equation}
\frac{d}{ds}\big(s^{-\lambda}\phi(s)\big)=\frac{1}{c_V}s^{-\lambda-1}r(s).
\end{equation}
Integrating from $s_0$ to $s\ge s_0$ and using $\phi(s_0)=\eta$ gives
\begin{equation}
s^{-\lambda}\phi(s) = s_0^{-\lambda}\eta + \frac{1}{c_V}\int_{s_0}^{s}\tau^{-\lambda-1}r(\tau)\,d\tau.
\end{equation}
Multiplying by $s^{\lambda}$ yields the claimed representation,
\begin{equation}
\phi(s)
=
\left(\frac{s}{s_0}\right)^{\lambda}\eta
+
\frac{1}{c_V}s^{\lambda}\int_{s_0}^{s}\tau^{-\lambda-1}r(\tau)\,d\tau.
\end{equation}

If $r\equiv 0$, the integral term vanishes and $\phi(s)=\eta (s/s_0)^{\lambda}$.

If additionally $r$ is linear, i.e. $r(s)=c_r s$ with $c_r\ge 0$, then
\begin{equation}
\int_{s_0}^{s}\tau^{-\lambda-1}r(\tau)\,d\tau
=
c_r\int_{s_0}^{s}\tau^{-\lambda}\,d\tau
=
\begin{cases}
\displaystyle \frac{c_r}{1-\lambda}\left(s^{1-\lambda}-s_0^{1-\lambda}\right), & \lambda\neq 1,\\[8pt]
\displaystyle c_r\log\!\left(\frac{s}{s_0}\right), & \lambda = 1.
\end{cases}
\end{equation}
Substituting into the general formula yields, for $\lambda\neq 1$,
\begin{equation}
\phi(s)
=
\eta\left(\frac{s}{s_0}\right)^{\lambda}
+
\frac{c_r}{c_V(1-\lambda)}\left(s - s^{\lambda}s_0^{1-\lambda}\right),
\end{equation}
and for $\lambda=1$,
\begin{equation}
\phi(s)
=
\eta\left(\frac{s}{s_0}\right)
+
\frac{c_r}{c_V}\,s\log\!\left(\frac{s}{s_0}\right).
\end{equation}
\end{proof}

\subsection*{E: Picking linear function $\phi$}
\begin{corollary_}
Let $\alpha_b(s)=c_b s$ and $\alpha_V(s)=c_V s$ with $c_b,c_V>0$. 
Assume $\phi(s)=c_\phi s$ for some constant $c_\phi\in\mathbb{R}$. 
Then $\phi$ satisfies
\[
\phi'(s)\alpha_V(s)=\alpha_b(\phi(s))+r(s), \qquad \forall s\ge 0,
\]
if and only if
\[
r(s)=c_\phi(c_V-c_b)\,s \qquad \forall s\ge 0.
\]
In particular, if $\phi\in\mathcal K$ and $r\in\mathcal K$, then necessarily $c_\phi>0$ and $c_V>c_b$,
and in that case $r(s)=c_r s$ with $c_r=c_\phi(c_V-c_b)>0$.
\end{corollary_}

\begin{proof}
If $\phi(s)=c_\phi s$, then $\phi'(s)=c_\phi$. Substituting into the ODE gives
\[
c_\phi(c_V s)=c_b(c_\phi s)+r(s),
\]
hence $r(s)=c_\phi(c_V-c_b)s$ for all $s\ge 0$. Conversely, if $r$ has this form, substitution shows
the identity holds for all $s\ge 0$. The final claim follows since $\phi\in\mathcal K$ implies $c_\phi>0$,
and $r\in\mathcal K$ requires $r(s)>0$ for $s>0$, i.e. $c_V-c_b>0$.
\end{proof}

\bibliographystyle{IEEEtran}
\bibliography{references}

@article{molnar2023safety,
  title={Safety-critical control with bounded inputs via reduced order models},
  author={Molnar, Tamas G and Ames, Aaron D},
  journal={arXiv preprint arXiv:2303.03247},
  year={2023}
}

@misc{bousias2025deepequivariantmultiagentcontrol,
      title={Deep Equivariant Multi-Agent Control Barrier Functions}, 
      author={Nikolaos Bousias and Lars Lindemann and George Pappas},
      year={2025},
      eprint={2506.07755},
      archivePrefix={arXiv},
      primaryClass={eess.SY},
      url={https://arxiv.org/abs/2506.07755}, 
}

@article{doi:10.1177/0278364911406761,
author = {Sertac Karaman and Emilio Frazzoli},
title ={Sampling-based algorithms for optimal motion planning},
journal = {The International Journal of Robotics Research},
volume = {30},
number = {7},
pages = {846-894},
year = {2011},
doi = {10.1177/0278364911406761},
URL = {https://doi.org/10.1177/0278364911406761},
eprint = {https://doi.org/10.1177/0278364911406761}
}

@Inbook{Richter2016,
author="Richter, Charles
and Bry, Adam
and Roy, Nicholas",
editor="Inaba, Masayuki
and Corke, Peter",
title="Polynomial Trajectory Planning for Aggressive Quadrotor Flight in Dense Indoor Environments",
bookTitle="Robotics Research: The 16th International Symposium ISRR",
year="2016",
publisher="Springer International Publishing",
address="Cham",
pages="649--666",
isbn="978-3-319-28872-7",
doi="10.1007/978-3-319-28872-7_37",
url="https://doi.org/10.1007/978-3-319-28872-7_37"
}

@INPROCEEDINGS{5980409,
  author={Mellinger, Daniel and Kumar, Vijay},
  booktitle={2011 IEEE International Conference on Robotics and Automation}, 
  title={Minimum snap trajectory generation and control for quadrotors}, 
  year={2011},
  volume={},
  number={},
  pages={2520-2525},
  doi={10.1109/ICRA.2011.5980409}}

@ARTICLE{9652122,
  author={Molnar, Tamas G. and Cosner, Ryan K. and Singletary, Andrew W. and Ubellacker, Wyatt and Ames, Aaron D.},
  journal={IEEE Robotics and Automation Letters}, 
  title={Model-Free Safety-Critical Control for Robotic Systems}, 
  year={2022},
  volume={7},
  number={2},
  pages={944-951},
  doi={10.1109/LRA.2021.3135569}}

@inproceedings{ames2019control,
  author    = {Ames, Aaron D. and Coogan, Samuel and Egerstedt, Magnus and Notomista, Gaetano and Sreenath, Koushil and Tabuada, Paulo},
  title     = {Control Barrier Functions: Theory and Applications},
  booktitle = {Proceedings of the European Control Conference (ECC)},
  year      = {2019},
  pages     = {3420--3431}
}

@article{ames2017control,
  author  = {Ames, Aaron D. and Xu, Xuan and Grizzle, J. W. and Tabuada, Paulo},
  title   = {Control Barrier Function Based Quadratic Programs for Safety Critical Systems},
  journal = {IEEE Transactions on Automatic Control},
  volume  = {62},
  number  = {8},
  pages   = {3861--3876},
  year    = {2017}
}

@article{girard2007approximation,
  author  = {Girard, Antoine and Pappas, George J.},
  title   = {Approximation Metrics for Discrete and Continuous Systems},
  journal = {IEEE Transactions on Automatic Control},
  volume  = {52},
  number  = {5},
  pages   = {782--798},
  year    = {2007}
}

@book{tabuada2009verification,
  author    = {Tabuada, Paulo},
  title     = {Verification and Control of Hybrid Systems: A Symbolic Approach},
  publisher = {Springer Science \& Business Media},
  year      = {2009}
}

@article{reissig2017feedback,
  author  = {Reissig, Gernot and Weber, Alexander and Rungger, Matthias},
  title   = {Feedback Refinement Relations for the Synthesis of Symbolic Controllers},
  journal = {IEEE Transactions on Automatic Control},
  volume  = {62},
  number  = {4},
  pages   = {1781--1796},
  year    = {2017}
}

@article{kloetzer2008fully,
  author  = {Kloetzer, Marius and Belta, Calin},
  title   = {A Fully Automated Framework for Control of Linear Systems from Temporal Logic Specifications},
  journal = {IEEE Transactions on Automatic Control},
  volume  = {53},
  number  = {1},
  pages   = {287--297},
  year    = {2008}
}

@article{rungger2017computing,
  author  = {Rungger, Matthias and Tabuada, Paulo},
  title   = {Computing Robust Controlled Invariant Sets of Linear Systems},
  journal = {IEEE Transactions on Automatic Control},
  volume  = {62},
  number  = {7},
  pages   = {3665--3670},
  year    = {2017}
}

@inproceedings{xiao2019control,
  author    = {Xiao, Wei and Belta, Calin},
  title     = {Control Barrier Functions for Systems with High Relative Degree},
  booktitle = {Proceedings of the IEEE Conference on Decision and Control (CDC)},
  year      = {2019},
  pages     = {474--479}
}

@inproceedings{nguyen2016exponential,
  author    = {Nguyen, Quan and Sreenath, Koushil},
  title     = {Exponential Control Barrier Functions for Enforcing High Relative-Degree Safety-Critical Constraints},
  booktitle = {Proceedings of the American Control Conference (ACC)},
  year      = {2016},
  pages     = {322--328}
}

@inproceedings{ames2014control,
  author    = {Ames, Aaron D. and Grizzle, J. W. and Tabuada, Paulo},
  title     = {Control Barrier Function Based Quadratic Programs with Application to Adaptive Cruise Control},
  booktitle = {Proceedings of the IEEE Conference on Decision and Control (CDC)},
  year      = {2014},
  pages     = {6271--6278}
}

@inproceedings{borrmann2015control,
  author    = {Borrmann, Ulf and Wang, Li and Ames, Aaron D. and Egerstedt, Magnus},
  title     = {Control Barrier Certificates for Safe Swarm Behavior},
  booktitle = {IFAC Workshop on Distributed Estimation and Control in Networked Systems},
  year      = {2015},
  pages     = {68--73}
}

@article{dawson2023safe,
  author  = {Dawson, Charles and Qin, Zhaoyang and Gao, Sicun and Fan, Chuchu},
  title   = {Safe Control with Learned Certificates: A Survey of Neural Lyapunov, Barrier, and Contraction Methods},
  journal = {IEEE Transactions on Automatic Control},
  year    = {2023}
}

@inproceedings{robey2020learning,
  author    = {Robey, Alexander and Hu, Hanwen and Lindemann, Lars and Zhang, Han and Dimarogonas, Dimos V. and Tu, Stephen and Matni, Nikolai},
  title     = {Learning Control Barrier Functions from Expert Demonstrations},
  booktitle = {Proceedings of the IEEE Conference on Decision and Control (CDC)},
  year      = {2020},
  pages     = {3717--3724}
}

\end{document}